\documentclass[prd,hidepacs,superscriptaddress,twocolumn,
floatfix,preprintnumbers,altaffilletter]{revtex4}
\usepackage{graphicx, url, amssymb, amsmath,longtable, rotating, color, units, wasysym, subfigure, epsfig, epstopdf, multirow, epstopdf, verbatim, appendix, lscape, rotating}

\newcommand{\aligo}{Advanced LIGO}

\begin{document}

\title{Validating gravitational-wave detections: \\ The Advanced LIGO hardware injection system}

\author{C.~Biwer}\affiliation{Syracuse University, Syracuse, NY 13244, USA}
\author{D.~Barker}\affiliation{LIGO Hanford Observatory, Richland, WA 99354, USA}
\author{J.~C.~Batch}\affiliation{LIGO Hanford Observatory, Richland, WA 99354, USA}
\author{J.~Betzwieser}\affiliation{LIGO Livingston Observatory, Livingston, LA 70754, USA}
\author{R.~P.~Fisher}\affiliation{Syracuse University, Syracuse, NY 13244, USA}
\author{E.~Goetz}\affiliation{LIGO Hanford Observatory, Richland, WA 99354, USA}
\author{S.~Kandhasamy}\affiliation{LIGO Livingston Observatory, Livingston, LA 70754, USA}
\author{S.~Karki}\affiliation{University of Oregon, Eugene, OR 97403, USA}
\author{J.~S.~Kissel}\affiliation{LIGO Hanford Observatory, Richland, WA 99354, USA}
\author{A.~P.~Lundgren}\affiliation{Albert-Einstein-Institut, Max-Planck-Institut f\"ur Gravitationsphysik, D-30167 Hannover, Germany}
\author{D.~M.~Macleod}\affiliation{LIGO Livingston Observatory, Livingston, LA 70754, USA}
\author{A.~Mullavey}\affiliation{LIGO Livingston Observatory, Livingston, LA 70754, USA}
\author{K.~Riles}\affiliation{University of Michigan, Ann Arbor, MI 48109, USA}
\author{J.~G.~Rollins}\affiliation{LIGO, California Institute of Technology, Pasadena, CA 91125, USA}
\author{K.~A.~Thorne}\affiliation{LIGO Livingston Observatory, Livingston, LA 70754, USA}
\author{E.~Thrane}\affiliation{Monash University, Victoria 3800, Australia}
\author{T.~D.~Abbott}\affiliation{Louisiana State University, Baton Rouge, LA 70803, USA}
\author{B.~Allen}\affiliation{Albert-Einstein-Institut, Max-Planck-Institut f\"ur Gravitationsphysik, D-30167 Hannover, Germany}\affiliation{University of Wisconsin-Milwaukee, Milwaukee, WI 53201, USA}\affiliation{Leibniz Universit{\"a}t Hannover, D-30167 Hannover, Germany}
\author{D.~A.~Brown}\affiliation{Syracuse University, Syracuse, NY 13244, USA}
\author{P.~Charlton}\affiliation{Charles Sturt University, Wagga Wagga, New South Wales 2678, Australia}
\author{S.~G.~Crowder}\affiliation{Bellevue College, Bellevue, WA 98008, USA}
\author{P.~Fritschel}\affiliation{LIGO, Massachusetts Institute of Technology, Cambridge, MA 02139, USA}
\author{J.~B.~Kanner}\affiliation{LIGO, California Institute of Technology, Pasadena, CA 91125, USA}
\author{M.~Landry}\affiliation{LIGO Hanford Observatory, Richland, WA 99354, USA}
\author{C.~Lazzaro}\affiliation{Center for Relativistic Astrophysics and School of Physics, Georgia Institute of Technology, Atlanta, GA 30332, USA}\affiliation{INFN, Sezione di Padova, I-35131 Padova, Italy}
\author{M.~Millhouse}\affiliation{Montana State University, Bozeman, MT 59717, USA}
\author{M.~Pitkin}\affiliation{SUPA, University of Glasgow, Glasgow, G12 8QQ, United Kingdom}
\author{R.~L.~Savage}\affiliation{LIGO Hanford Observatory, Richland, WA 99354, USA}
\author{P.~Shawhan}\affiliation{University of Maryland, College Park, MD 20742, USA}
\author{D.~H.~Shoemaker}\affiliation{LIGO, Massachusetts Institute of Technology, Cambridge, MA 02139, USA}
\author{J.~R.~Smith}\affiliation{California State University Fullerton, Fullerton, CA 92831, USA}
\author{L.~Sun}\affiliation{The University of Melbourne, Parkville, Victoria 3010, Australia}
\author{J.~Veitch}\affiliation{University of Birmingham, Birmingham, B15 2TT, United Kingdom}
\author{S.~Vitale}\affiliation{LIGO, Massachusetts Institute of Technology, Cambridge, MA 02139, USA}
\author{A.~J.~Weinstein}\affiliation{LIGO, California Institute of Technology, Pasadena, CA 91125, USA}
\author{N.~Cornish}\affiliation{Montana State University, Bozeman, MT 59717, USA}
\author{R.~C.~Essick}\affiliation{LIGO, Massachusetts Institute of Technology, Cambridge, MA 02139, USA}
\author{M.~Fays}\affiliation{Cardiff University, Cardiff CF24 3AA, United Kingdom}
\author{E.~Katsavounidis}\affiliation{LIGO, Massachusetts Institute of Technology, Cambridge, MA 02139, USA}
\author{J.~Lange}\affiliation{Rochester Institute of Technology, Rochester, NY 14623, USA}
\author{T.~B.~Littenberg}\affiliation{University of Alabama in Huntsville, Huntsville, AL 35899, USA}
\author{R.~Lynch}\affiliation{LIGO, Massachusetts Institute of Technology, Cambridge, MA 02139, USA}
\author{P.~M.~Meyers}\affiliation{University of Minnesota, Minneapolis, MN 55455, USA}
\author{F.~Pannarale}\affiliation{Cardiff University, Cardiff CF24 3AA, United Kingdom}
\author{R.~Prix}\affiliation{Albert-Einstein-Institut, Max-Planck-Institut f\"ur Gravitationsphysik, D-30167 Hannover, Germany}
\author{R.~O'Shaughnessy}\affiliation{Rochester Institute of Technology, Rochester, NY 14623, USA}
\author{D. Sigg}\affiliation{LIGO Hanford Observatory, Richland, WA 99354, USA}

\begin{abstract}
Hardware injections are simulated gravitational-wave signals added to the Laser Interferometer Gravitational-wave Observatory (LIGO).
The detectors' test masses are physically displaced by an actuator in order to simulate the effects of a gravitational wave. The simulated signal initiates a control-system response which mimics that of a true gravitational wave.
This provides an end-to-end test of LIGO's ability to observe gravitational waves.
The gravitational-wave analyses used to detect and characterize signals are exercised with hardware injections.
By looking for discrepancies between the injected and recovered signals, we are able to characterize the performance of analyses and the coupling of instrumental subsystems to the detectors' output channels.
This paper describes the hardware injection system and the recovery of injected signals representing binary black hole mergers, a stochastic gravitational wave background, spinning neutron stars, and sine-Gaussians.
\end{abstract}

\maketitle

\section{Introduction}\label{introduction}

The Advanced Laser Interferometer Gravitational-Wave Observatory (\aligo ) is a network of two interferometric gravitational-wave detectors located in Hanford, WA, and Livingston, LA~\cite{aligo}.
The \aligo\ detectors are part of a global network of current and planned detectors including Virgo~\cite{advirgo}, GEO600~\cite{geo}, KAGRA~\cite{kagra}, and LIGO India~\cite{LIGO_India}.
The first direct observations of gravitational waves, both from binary black hole mergers, were made in \aligo's first observing run~\cite{gw150914_detection,gw151226_detection}.

In order to to make confident statements about gravitational-wave events, \aligo\ employs studies to understand both transient and persistent noise artifacts~\cite{DetcharGW150914}, and the calibration of the detectors~\cite{S6detchar,gw150914_calibration}.
In addition, analysis pipelines for detection and parameter estimation of gravitational-wave signals employ a wide range of different techniques to mitigate the impact of non-Gaussian and non-stationary noise in the detectors' data.
Testing these analyses and characterizing the detectors involves carrying out ``hardware injections'' in which we simulate the detectors' response to a gravitational-wave signal.
Hardware injections are used to validate ``software injections'', where simulated signals are added to the data as part of an analysis pipeline without any physical actuation occurring; software injections are used for high-statistics evaluation of the performance of analyses.

Hardware injections have several other uses.
Following a detection candidate, we study similar simulated gravitational-wave signals through the use of repeated injections. These hardware injections provide an end-to-end check for the search and parameter estimation analyses to recover signals in the detectors' data.
The recovery of hardware injections provides an additional check of the sign of the calibration between the \aligo\ detectors using astrophysical waveforms and the recovery measures the time delay of the signal in the controls system; the calibration of the detectors is checked by other means as well~\cite{S6detchar,gw150914_calibration}.
In addition, we can check for instrumental and environmental channels that respond to changes in differential arm length variations from gravitational-wave signals.

Another use for hardware injection in Initial LIGO were ``blind injections'' which were hardware injections known only to a small team~\cite{blind,Abadie:2010mt}. Blind injections simulate the detection and characterization of a real astrophysical signal.
No blind injections were carried out during \aligo's first observing run. There are no plans to perform blind injections in future observing runs.

To create a hardware injection we physically displace the detectors' test masses.
The mirrors in the arms act as ``freely falling'' test masses~\cite{gw150914_detector}.
\aligo\ measures the differential displacement along the two arms ${\Delta L = L_x - L_y}$, and the output channel to analyses is gravitational-wave strain ${h = \Delta L / L}$ where ${L = (L_x + L_y) / 2}$~\cite{gw150914_detector}.
Differential displacement of the test masses mimics the detectors' response to a gravitational-wave signal.

The detectors' response to a true gravitational-wave is not exactly the same as the detectors' response to physically displacing the test masses~\cite{thesis_kawabe,high_freq_response}.
However, the difference is well understood, and it is only relevant at high frequencies~\cite{thesis_kawabe,high_freq_response}.
In addition, the actuators apply a force to the test masses in their suspensions whereas a true gravitational-wave does not.

Advanced LIGO uses different actuators to perform hardware injections than Initial LIGO.
In Initial LIGO, the test masses were displaced using magnets mounted on the the optic itself, however, these actuators are no longer used to move the test masses due to displacement noise~\cite{barkhausen,Smith:2009bx}.
In \aligo's first observing run, hardware injections were realized with two different actuation methods: electrostatic drive systems~\cite{sensors_actuators} and photon radiation pressure actuators referred to as ``photon calibrators''~\cite{2016arXiv160805055K}.
Starting in December 2015 the photon calibrators have been the only actuator used to perform hardware injections since their actuation range available for hardware injections is larger.

During \aligo's first observing run a wide variety of waveforms were injected.
\aligo\ is sensitive to astrophysical sources of gravitational waves including: binary black hole and/or neutron star mergers~\cite{Abbott:2016ymx,O1BBH}, the stochastic gravitational-wave background~\cite{2000PhR...331..283M}, and spinning neutron stars~\cite{TargetedCWSearches}.
Hardware injections for each of these astrophysical sources were performed.
In addition, detector characterization studies injected series of sine-Gaussians across the \aligo\ frequency range.

This paper describes how we inject signals into the \aligo\ detectors with the photon calibrators in Section~\ref{infrastructure}. Section~\ref{recovery} describes the results from analyses that used hardware injections in \aligo's first observing run.
This includes the recovery of binary black hole merger signals in Section~\ref{cbc} and \ref{sec:burst}, the stochastic gravitational-wave hardware injection in Section~\ref{sec:stochastic}, and a population of spinning neutron stars in Section~\ref{sec:cw}.
Although binary neutron star hardware injections were performed we do not discuss their recoveries in this paper.
A description of the detector characterization analysis to check the response of instrumental and environmental changes to differential displacement of the test masses is described in Section~\ref{sec:detchar}. Finally, Section~\ref{sec:conclusions} summarizes the hardware injections from \aligo's first observing run.

\section{Hardware Injection Procedure}\label{infrastructure}

Each different type of astrophysical source has different signal characteristics and properties, and hence different technical requirements for the hardware injection system.
In particular, the difference in the time duration of the sources.

The time in \aligo's frequency range for compact-object binary mergers can last a fraction of a second to minutes depending on the component masses.
The signal enters \aligo's frequency range from low frequency, and as the two component masses inspiral closer together they sweep upward in frequency~\cite{gw150914_detection}.
The merger's termination frequency and waveform length is determined by the masses of the two objects.
For example, GW150914 terminates at $250$~Hz after about 0.2~s above 35~Hz~\cite{gw150914_detection}, whereas the inspiral-only portion of a binary neutron star waveform with both component masses equal to $1.4$~M$_\odot$ terminates at $1527$~Hz after about 36~s above 35~Hz.

Gravitational-waves from a spinning neutron star will be present in the data for the full duration of an observing run.
Spinning neutron stars emit continuous gravitational waves at an almost constant frequency which is Doppler modulated by Earth's motion~\cite{TargetedCWSearches}.
The gravitational-wave frequency slowly evolves as the pulsar spins down~\cite{TargetedCWSearches}.

The stochastic gravitational-wave background will persist in the data throughout the observing run.
The stochastic background is the superposition of many events that combine to create a low-level broadband non-deterministic signal~\cite{2000PhR...331..283M}.

Non-astrophysically motivated injections for detector characterization studies use a succession of short duration ($<1$~s) sine-Gaussians across \aligo's frequency range.

We categorize hardware injections into two classes: ``transient injections'' that are localized in time, and ``continuous-wave injections'' that are active throughout the duration of the observing run.
Examples of transient injections include simulated binary black hole and/or neutron star mergers, sine-Gaussians, and stochastic background signals.
These signals have a finite duration.
The simulated stochastic background is included as a transient injection since we increase the amplitude of the waveform in order to limit it to a short segment of data.
Continuous-wave injections simulate a synthetic population of rapidly spinning neutron stars (which we designate in shorthand as pulsars, although such a source need not emit electromagnetic pulsations detectable at the Earth).

\begin{figure*}[t]
\centering
\includegraphics[width=\linewidth]{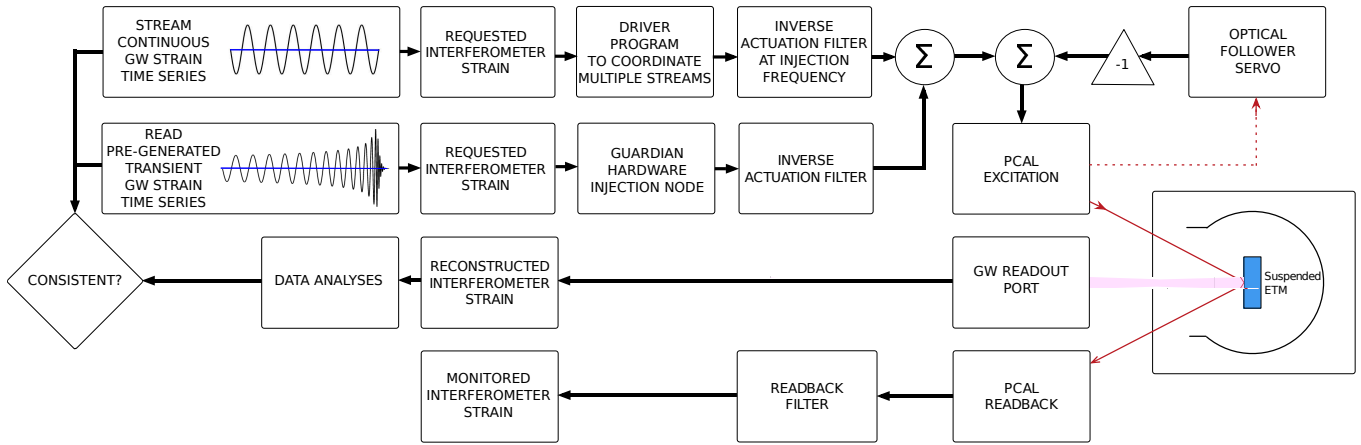}
\caption{
Block overview of the \aligo\ hardware injection system. Time series for transient and continuous-wave injections are generated and sent to the photon calibrator (PCAL). The signal modulates the laser power of the photon calibrator to displace the end test mass (ETM) in a way that mimics a gravitational wave (GW) passing through the detector. The optical follower servo has its own pick-off of the light that is sent towards the ETM indicated by a dashed line. The gravitational-wave strain of the detector is analyzed and checked for consistency by the analysis' developers. A photodetector that receives the light reflected from the test mass is used to monitor and verify the injected signal.
}
\label{fig:software}
\end{figure*}

Separate automation processes control transient and continuous-wave injections. 
Fig.~\ref{fig:software} shows a schematic of the two pathways that generate and transmit gravitational-wave strain time series to the photon calibrator.
In this section we work through Fig.~\ref{fig:software}, beginning at the top-left and working clockwise, in order to describe the processes that control the transient and continuous-wave injections.

We generate the simulated gravitational waveforms for transient injection signals prior to injection.
The system for managing the automated processes of the \aligo\ detector subsystems is Guardian~\cite{guardian}.
Guardian manages the transient hardware injections, it reads the next scheduled injection's time series and transmits the data to the digital control system of the detector at the scheduled time.

Continuous-wave injections are generated in real-time. A streaming time series of simulated gravitational waves from a synthetic population of spinning neutron stars, described by astrophysical parameters, including the strain amplitude, sky location, and initial frequency, is transmitted to the digital controls system of the detector.
A driver program called \texttt{psinject} (``pulsar injection'')
coordinates the simultaneous generation and buffering of multiple streams of signals representing pulsars~\cite{lalsuite}.

The transient and continuous-wave signals in the digital controls system of the detector are sent to an actuator that displaces the test masses to simulate the detector's response to a gravitational wave signal.
In \aligo's first observing run, we used the electrostatic drive systems~\cite{sensors_actuators} and photon calibrators~\cite{2016arXiv160805055K} as actuators for hardware injections.
Each actuator has its own actuation pathway in the controls system; however, in Fig.~\ref{fig:software} we show only the photon calibrators' pathway.

Hardware injections are carried out by actuating one of the end test masses (ETM) of the interferometer and thus inducing differential interferometer strain variations that simulate the response to an incident gravitational wave.
We only need to apply a force on one ETM to induce differential strain variations in the interferometer.
The common arm length degree of freedom of
the interferometer, $(L_x + L_y) / 2$, is controlled by its own servo.
If an actuator lengthens the $x$-arm by applying a force on the ETM, then the common arm length servo will promptly shorten the $y$-arm length to suppress the change in the common arm length degree of freedom.
This creates differential interferometer strain variations that are partially suppressed by the differential arm length feedback servo.

The differential arm length degree of freedom of the interferometer is controlled by a feedback servo loop that actuates the longitudinal position of one of the ETMs~\cite{gw150914_calibration}.
The differential arm length feedback control loop suppresses apparent ETM displacements resulting from noise sources, signal injections, and gravitational waves.
Because this servo suppresses the injected waveform, reconstructing the unsuppressed injected strain requires correcting for the action of this servo.
The correction for the response of the common and differential arm length servos used in reconstructing the gravitational-wave strain is described in~\cite{gw150914_calibration}.

The actuators for the servo that controls the differential arm length degree of freedom are electrostatic drive systems. These actuators apply forces via fringing field gradients from electrodes patterned onto a reaction mass separated by a few millimeters from the back surface of the ETM ~\cite{sensors_actuators}.

The electrostatic drive systems were used at the beginning of \aligo's first observing run for injecting simulated signals.
They successfully injected the waveforms for the GW150914 and stochastic background hardware injection analyses.
However, the actuation range available for hardware injections is restricted because they are part of the differential arm length servo which consumes a significant fraction of its total actuation range in maintaining stable servo operation.

In order to inject a larger parameter space of waveforms, for example binary black hole and/or neutron star mergers at closer distances, we transitioned to photon calibrators for hardware injections.
Since December 2015, we use a photon calibration system to displace the ETM in a way that simulates the effect of a gravitational wave signal. This is depicted on the right of Fig.~\ref{fig:software}.

A photon calibrator system uses an auxiliary, power-modulated laser with two beams impinging on the ETM located at the end of the $x$-arm of the interferometer.
The photon calibrator on the other arm, the $y$-arm, is used for calibrating the detector output~\cite{2016arXiv160805055K}. The two beams are diametrically opposed on the surface of the ETM, adjusted to have equal powers, and positioned to minimize unintended torques and deformations of the surface which could cause errors in the expected displacement.

The \aligo\ photon calibrators employ a feedback control system referred to as the ``optical follower servo''~\cite{2016arXiv160805055K,ofs}.  This servo, with a bandwidth of $\sim$100~kHz, facilitates simulated signal injection via ETM actuation. This ensures that the laser output power modulation closely follows the analog voltage waveform injected at the servo input.

Digital infinite impulse response (IIR) compensation filters, called the ``inverse actuation filters,'' convert the requested interferometer strain signal (a digital signal) into an estimate of the photon calibrator optical follower servo input signal (an analog signal) required to achieve the desired length actuation.
There is an analogous set of filters for the electrostatic drive system; however, we focus on the photon calibrators here.
These filters are designed to compensate for several factors.
There is compensation for: (i) the force-to-length transfer function of the suspended ETM, 
(ii) the signal conditioning electronics that includes a digital anti-imaging filter, the digital to analog converter gain, and an analog anti-imaging filter, and (iii) the optical follower servo transfer function.
Phase delays of the anti-imaging filters and physical time delays of the digital control system cannot be compensated by the inverse actuation filters because the digital IIR filters allowed by the \aligo\ control system must be causal.
Thus, injections have known, but uncompensated, delays that we take into account during injection recovery.

The digital signals from the transient and continuous-wave injection pathways are passed through the inverse actuation filters, summed, and sent to the photon calibrator; see Fig.~\ref{fig:software}.
Sporadic, unintended interruptions occurred in the Hanford injection system during \aligo's first observing run, in which the buffering failed to keep up with real-time injection.
The cause was not tracked down because the interruptions occurred at apparently random times, but the drop-outs may be related to periods of high traffic on the controls system computer network.
The sudden termination introduces a step function to the inverse actuation filters that has a large response at high frequencies.
The effect of these  dropouts, should they recur, will be mitigated by the use of point-by-point, Fourier-domain inverse actuation functions, using a separate, constant coefficient for each of the injected spinning neutron stars, all of which are extremely narrowband. This is shown in the continuous-wave injection pathway in Fig.~\ref{fig:software}.
Transient injections were not affected.
Guardian sets the gain after the inverse actuation filters to zero while there is no active transient injection so unintended signals do not propagate into the detector data.

The strain actually injected into the interferometer is determined using the photon calibrator read-back signal generated by a power sensor that monitors the laser light reflected from the ETM, as shown at the bottom of Fig.~\ref{fig:software}.  The output of this sensor is converted to injected interferometer strain using the read-back filter that compensates for the force-to-length transfer function as well as digital and analog filters in the signal read-back pathway. In the case of hardware injections, however, the excitation channel is calibrated by taking a transfer function measurement between the excitation channel and the read-back photodetectors.
This transfer function is then incorporated within the inverse actuation filters. This provides a calibration accuracy on the order of a few percent, sufficient for the hardware injection analysis. For better calibration, however, we can compare the recovered signal and the injected signal as measured by the read-back photodetector.

\begin{figure}[t]
\centering
\includegraphics[width=\columnwidth]{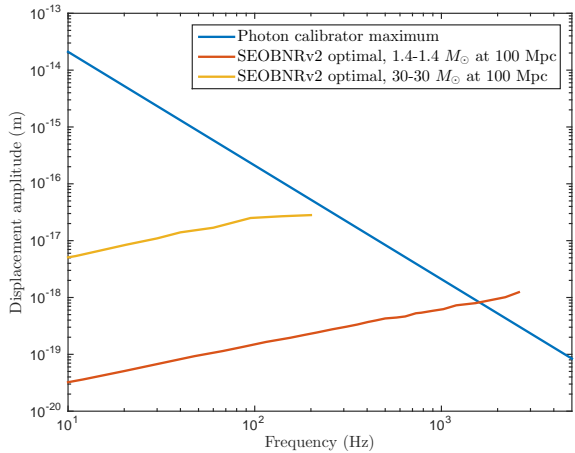}
\caption{
The maximum displacement of an ETM using the photon calibrator (blue). For the sinusoidal force induced by sinusoidally power modulated laser beams, $F= mA$ implies that the induced displacemnet is given by $x=-F/(m \omega)^2$. The maximum displacement of the ETM required for two optimally-oriented compact binary waveforms that contain an inspiral, merger, and ringdown are shown for reference. A 3-30~M${}_{\odot}$ binary at 100~Mpc (yellow) and 1.4-1.4~M${}_{\odot}$ binary at 100~Mpc (red) were generated using the SEOBNRv2 approximant~\cite{seobnrv2}. Note that the required displacement for the 1.4-1.4~M${}_{\odot}$ binary exceeds the maximal photon calibrator displacement at high frequencies.
}
\label{fig:pcal_range}
\end{figure}

There are some limitations to the photon calibrator system. 
First, the photon calibrator has a limited actuation strength. Fig.~\ref{fig:pcal_range} shows the maximal displacement of the ETM using the photon calibrator system. The photon calibrator can provide up to $\sim$1~W of peak power, but the force-to-length response of the ETM transfer function scales as the inverse-square of frequency~\cite{iLIGOpcal}.
Thus, the photon calibrator is limited in the amount of induced ETM displacement, especially at higher frequencies.
Second, signal fidelity above $\sim$1~kHz is limited due to the shape of the anti-imaging filters and the desire to roll off the compensation filters close to the Nyquist frequency such that the compensation filters are not unstable.
Nonetheless, the photon calibrator is able to provide precise, calibrated displacements of the ETM in response to many astrophysical waveforms.

\section{Results}\label{recovery}

In this section we describe results from hardware injection analyses in \aligo's first observing run. Hardware injections were used as: (i) an end-to-end test of searches and parameter estimation analyses, (ii) an additional check of the calibration, and (iii) a method to check for instrumental and environmental channels that are coupled to the detectors' output channels.
Binary black hole waveforms with parameters similar to GW150914 and GW151225 were used to test modeled and unmodeled analyses described in Section~\ref{cbc} and \ref{sec:burst}. A simulated stochastic gravitational-wave signal was recovered with the analysis described in Section~\ref{sec:stochastic}. The population of simulated spinning neutron stars analysis in Section~\ref{sec:cw} also provided an additional check of the calibration; they were used to verify the overall sign of the detectors' calibration and to measure the time delay between the hardware injection excitation channel and the detectors' output channels.
It is possible to use other hardware injections to perform these checks of the calibration as well.
These astrophysical signals were injected coherently into the Hanford (H1) and Livingston (L1) detectors.
Section~\ref{sec:detchar} describes the study that injects a series of sine-Gaussians across the \aligo\ frequency range to check for instrumental and environmental channels that respond to differential arm length variations.

\subsection{Compact Binary Coalescence Gravitational-Wave Hardware Injections}\label{cbc}

\aligo\ observed two binary black hole mergers (GW150914 and GW151226) and a third detection candidate (LVT151012) during its first observing run~\cite{gw150914_detection,gw151226_detection}.
After each detection was made, hardware injections were used to simulate gravitational-wave sources with similar parameters to each event in the detector. Verifying that these hardware injections were recovered by the search and parameter estimation analyses was part of the validation of each detection.
Compact binary coalescence searches use matched filtering to correlate \aligo\ data with a bank of gravitational-wave templates~\cite{Allen:2005fk}.
Here we consider hardware injections analyzed by the PyCBC search for gravitational waves~\cite{pycbc,alex_nitz_2016_197080} described in~\cite{gw150914_cbc,O1BBH}.
Parameter estimation analyses were used to analyze the hardware injections and check for consistency with GW150914 and GW151226. We ran the same code used to characterize the detected events~\cite{Veitch:2014wba,gw150914_pe}.
We show the recovery of hardware injections with parameters taken from posterior distributions of parameter estimation results for GW150914~\cite{gw150914_pe} and GW151226~\cite{gw151226_detection,O1BBH}.

For GW150914 and GW151226, we injected ten waveforms coherently into the two detectors after collecting enough data to confidently establish a detection.
The GW150914 hardware injections were generated with the SEOBNRv2 waveform approximant and included systems with component spins aligned with the angular momentum of the binary~\cite{seobnrv2}.
These signals were injected October 2 to October 6, 2015.
The GW151226 hardware injections were generated with the precessing waveform approximant IMRPhenomPv2~\cite{imrphenom,imrphenom2} and injected on January 11, 2016.

\begin{figure}[t]
\centering
\includegraphics[width=\columnwidth]{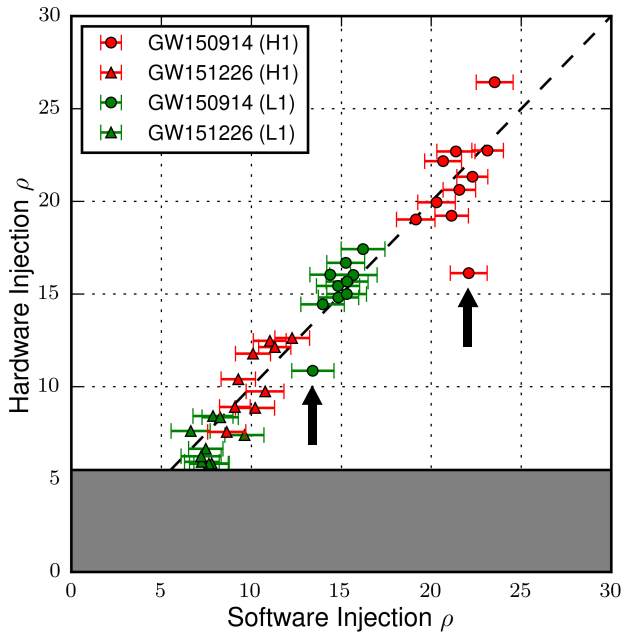}
\caption{A comparison of the signal-to-noise ratio $\rho$ from software injections and the recovered signal-to-noise ratio of the hardware injection. Parameters for the hardware injections were drawn from the posterior distributions for GW150914 (circles) and GW151226 (triangles). The software injection $\rho$ is the mean and $1\sigma$ error from the recovery of 50 software injections filtered with the injected waveform near the time of the injection. The threshold on $\rho$ is indicated by the gray region. The arrows indicate the coherent injection affected by a nearby noise transient.}
\label{fig:pycbc_rec}
\end{figure}

Fig~\ref{fig:pycbc_rec} shows the reported PyCBC matched-filter signal-to-noise ratio $\rho$ versus the expected $\rho$ calculated using software injections.
Fig.~\ref{fig:pycbc_rec} includes 19 of the 20 hardware injections performed for GW150914 and GW151226.
The expected $\rho$ was calculated using software injections, in which signals are added to the data without any physical actuation.
The normalization of $\rho$ implies that the $\rho$ measured for a population of identical signals in different realizations of the detector noise will be $\int df |\tilde h(f)|^2 / S_h(f)$~\cite{Allen:2005fk}.
The recovered software injections were found to be consistent with the expectation.

All of the hardware injections are coherent but an astrophysical signal can have a different $\rho$ in each detector. Hanford and Livingston have their own angular sensitivity and noise spectra that affects $\rho$ for an event~\cite{gw150914_detector}.
One of the GW150914 hardware injections reported $\rho<5.5$ in Livingston.
In order to manage computational considerations, the analysis requires a single-detector signal-to-noise ratio of at least 5.5.
Thus, this injection was not ``detected.''
A signal-to-noise ratio $<5.5$ for this injection, with an expected signal-to-noise ratio of $6$, is consistent with the variation of the matched-filter output in Gaussian noise~\cite{Allen:2005fk}.

In Fig.~\ref{fig:pycbc_rec} there is one GW150914 hardware injection that was recovered with a signal-to-noise ratio of 16.1 and 10.9 in Hanford and Livingston respectively; however, the injection had an expected signal-to-noise ratio of 22.1 and 13.4. This injection was recovered with a lower signal-to-noise ratio because a loud transient noise artifact was present in the Livingston data shortly after the hardware injection.

There are a variety of transient noise artifacts that adversely affect the search~\cite{DetcharGW150914,S6detchar}.
The search includes a signal consistency test to mitigate their effect.
PyCBC reports a detection statistic $\hat{\rho}$~\cite{newsnr} that combines information about the matched-filter signal-to-noise ratio $\rho$, where consistency is determined with a reduced $\chi^2$ statistic $\chi^{2}_{r}$~\cite{chisquare}. 
The $\chi^{2}_{r}$ statistic~\cite{chisquare} downweights the significance of the noise transients.

While hardware injections are an important end-to-end test, software injections are useful because a large number can be performed without disturbing the detector or significantly reducing the duty cycle of the detectors.
Fig.~\ref{fig:pycbc_rec} shows the software injections to be consistent with the recovery of signals that propagate through the detectors, therefore we can generate large populations of software injections that are used in other studies to evaluate the search efficiency~\cite{pycbc}, detections~\cite{gw150914_cbc}, and binary merger rates~\cite{O1BBH,Abbott:2016ymx}.

\begin{figure*}[t]
\centering
\includegraphics{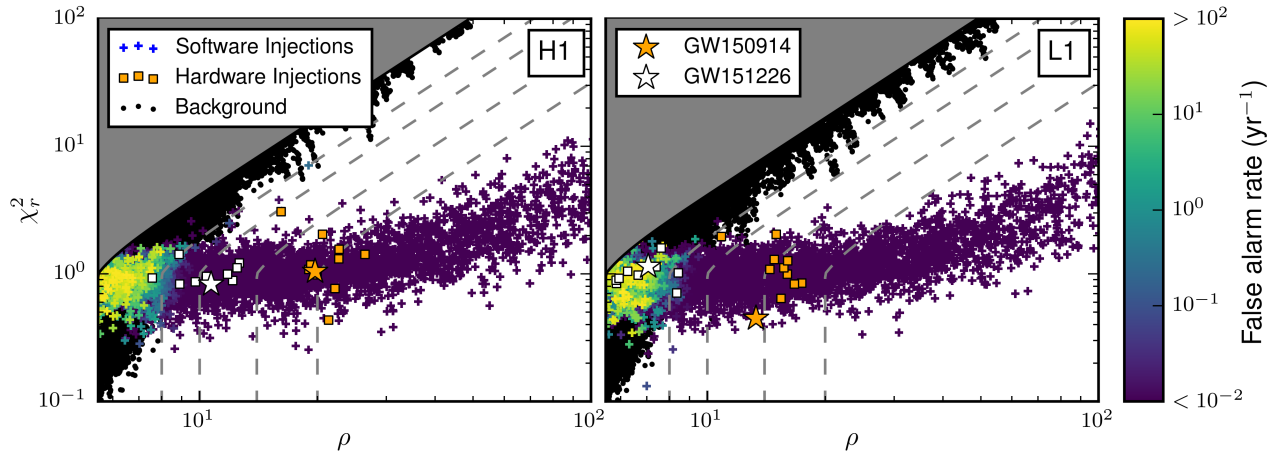}
\caption{PyCBC $\chi^{2}_{r}$ statistic versus matched-filter signal-to-noise ratio $\rho$ for each detector. Software injections are represented as pluses that are colored by false-alarm rate.
The false-alarm rate is calculated using the time-slide algorithm described in~\cite{pycbc}.
The gravitational-wave events GW150914 and GW151226 are shown as stars. Hardware injections for GW150914 and GW151226 are represented as boxes. These are coherent software and hardware injections, therefore the H1 and L1 plots are dependent on each other. Single-detector background distributions (black dots) are plotted; there was a threshold applied indicated by the gray region. Lines of constant detection statistic $\hat\rho$ are shown (gray dashed lines); plotted are $\hat{\rho} = \{8, 10, 14, 20\}$.
}
\label{fig:cbc}
\end{figure*}

Fig.~\ref{fig:cbc} shows the $\chi^{2}_{r}$ statistic~\cite{chisquare} versus $\rho$ for hardware injections, a large population of software injections, and noise transients.
Astrophysical events are indicated with stars.
Hardware injections are indicated with squares.
Software injections are denoted by pluses.
These software simulations repeat the analysis many times to test the search across a large parameter space.
The software injections in Fig.~\ref{fig:cbc} were generated from a population of aligned-spin binaries with source-frame component masses between $2$ to $98~$M${}_\odot$ using the SEOBNRv2 waveform approximant~\cite{seobnrv2}.
The population of software injections is randomly distributed in sky location, orientation, distance, and time. The injection times are within the 39~day period around GW150914 reported in~\cite{gw150914_cbc}.

In Fig.~\ref{fig:cbc} a highly significant astrophysical signal should be clearly separated from the background distributions.
We see a separation of the software injections with high significance (false-alarm rate $<\unit[1/100]{yr^{-1}}$) and background distributions.
All ten GW150914 hardware injections are recovered with high significance.
Although the GW151226 Livingston hardware injections are not visibly distinguishable from the background distribution in Fig.~\ref{fig:cbc}, seven hardware injections have a highly significant false-alarm rate ($<\unit[1/100]{yr^{-1}}$) since we combine data from both detectors.
Two hardware injections were recovered with $\unit[1/10]{yr^{-1}}>$ false-alarm rate $>\unit[1/100]{yr^{-1}}$, a significance comparable to the gravitational-wave candidate LVT151012 ($\unit[1/2]{yr^{-1}}$) reported in \aligo's first observing run~\cite{gw150914_detection}.
The software and hardware injections with similar parameters to GW150914 and GW151226 found with high significance validates the search's ability to detect similar systems.

If a detection candidate is a true gravitational wave, we should be able to reproduce the morphology of the posterior distributions using the hardware injections as well as with software injection.
Conversely any significant differences have the potential to highlight discrepancies between the observation and our waveform models, or errors in our data analysis.
Here we focus on two parameters: chirp mass and sky location.

The chirp mass $\cal{M}$ is defined as
\begin{equation} 
  {\cal M} = \frac{{(m_1 m_2)}^{3/5}}{{(m_1+ m_2)}^{1/5}} .\label{eqn:mchirp}
\end{equation}
Here, $m_1$ and $m_2$ are the binary's component masses. The chirp mass is typically the best estimated parameter of a compact binary coalescence signal, since it dominates the phase evolution during inspiral. 
In Fig.~\ref{fig:mc} we show for all the GW151226 hardware injections the posterior distributions of the chirp mass minus the respective injected values, using the precessing waveform approximant IMRPhenomPv2~\cite{imrphenom,imrphenom2}. Most posteriors have comparable width.
Hardware injections with low signal-to-noise ratio have broader distributions and in one case shows bimodality.
The width of the 90\% credible interval for the detector-frame chirp mass for GW151226 is $\sim0.12~$M${}_{\odot}$~\cite{O1BBH}, which is comparable to that found with the hardware injections.
Verifying that the width and shape of the posterior distribution for the chirp mass of the candidate events is similar to those of the hardware injection analyses has been part of validating the parameter estimation results for each detection.

\begin{figure*}[t]
\centering
\includegraphics[width=1.9\columnwidth]{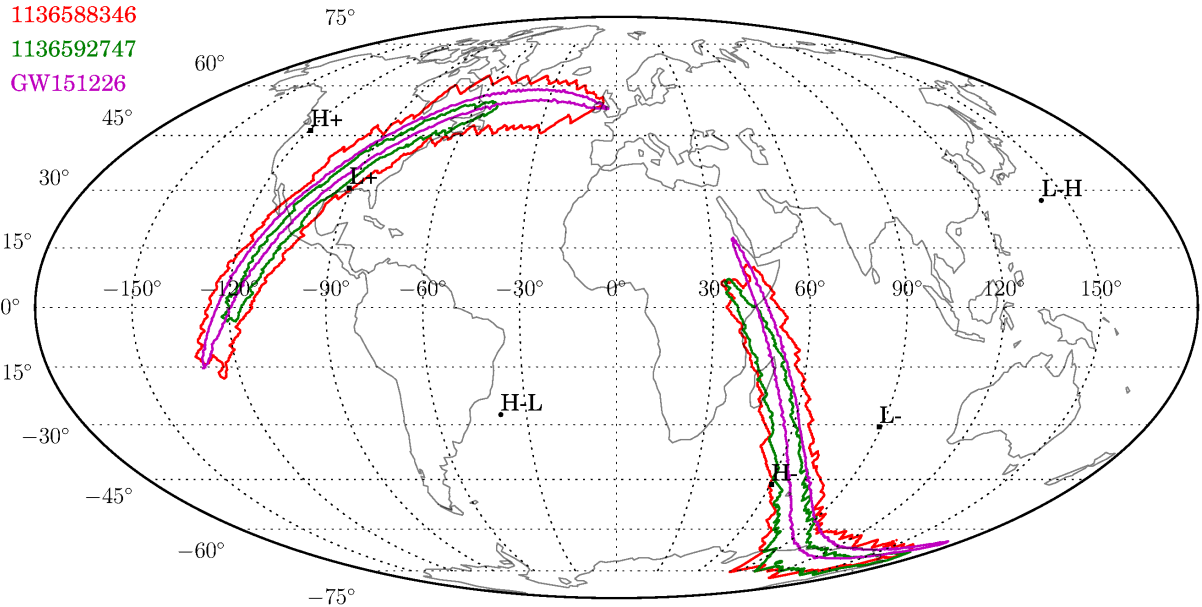}
\caption{The 90\% confidence interval skymaps for two hardware injection (red and green) and GW151226 (magenta). The skymaps are shown in Earth-bound coordinates. H+ and L+ mark the Hanford and Livingston sites, and H- and L- indicate antipodal points; H-L and L-H mark the poles of the line connecting the two detectors (the points of maximal time delay). The two hardware injections are chosen to be representative of an average event (green) and a sub-threshold event (red). We notice how all sky maps have support near the same ring of equal time delay between the two \aligo\ detectors.}
\label{fig:sky_gw151226}
\end{figure*}

\begin{figure}[t]
\includegraphics[width=\columnwidth]{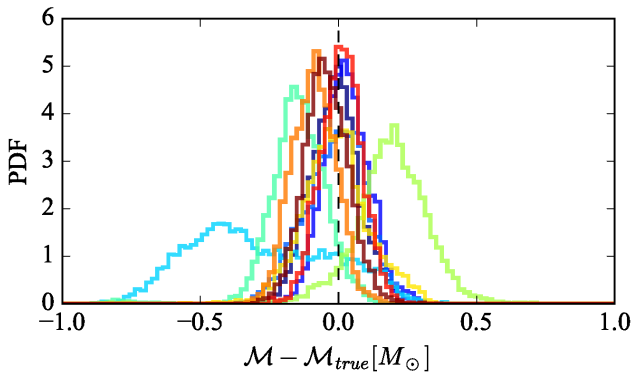}
\caption{Posterior probability density functions (PDF) for the chirp mass inferred from GW151226 hardware injections. The true value has been removed to center all distributions around zero. Hardware injections with very low signal-to-noise ratio show large width and in one case bimodality. The bimodal distribution comes from the injection at GPS time 1136588346 which is also shown in Fig.~\ref{fig:sky_gw151226}.}
\label{fig:mc}
\end{figure}

Sky maps from the parameter estimation analysis of GW150914 and GW151226 were shared with electromagnetic observatories~\cite{gw150914_gcn,gw151226_gcn} and are shown in~\cite{O1BBH,losc_gw151226}. 
In Fig.~\ref{fig:sky_gw151226}, we show a reconstructed Earth-bound coordinate sky map for GW151226 along with sky maps for two hardware injections.
One of the two hardware injections (at GPS time 1136588346) has low signal-to-noise ratio and thus spans a larger sky area, although still near to the same triangulation ring.
The other injection (at GPS time 136592747) is instead representative of the typical map: all other maps look similar to this and are not shown to avoid overcrowding. 

A previous study used the parameter estimation method described above to validate another strategy used to interpret GW150914, by directly comparing data to simulations of Einstein's equations~\cite{NRPaper}.
In that study the parameter estimates for GW150914 derived from IMRPhenomPv2 and numerical relativity agreed~\cite{NRPaper}.  
This study was repeated comparing hardware injections to numerical relativity, and found posterior distributions in mass and spin that were consistent with the IMRPhenomPv2 analysis.
This comparison provided a timely, independent validation of a this new parameter estimation strategy using real data and in this region of parameter space.
\subsection{Burst Gravitational-Wave Hardware Injections}\label{sec:burst}
There are astrophysical sources of gravitational waves that have poorly modeled or unknown waveforms, such as
core-collapse supernovae~\cite{living_review}.
In order to search for a wide range of unmodeled astrophysical sources, analyses search the \aligo\ strain data for short-duration, transient gravitational-wave events referred to as ``bursts''~\cite{gw150914_burst}.
Here, we look at injection recoveries using: Coherent WaveBurst~\cite{cwb}, BayesWave~\cite{BayesWave}, and LALInferenceBurst~\cite{Veitch:2014wba,skilling_nest}.
These analyses produce reconstructed waveforms with minimal assumptions about the waveform morphology. We compare these reconstructions to the injected waveforms of hardware injections.

Binary black hole waveforms were used to test the burst analyses.
In addition to the ten GW150914 hardware injections described in Section~\ref{cbc}, there were 24 waveforms injected with physical parameters similar to GW150914.  Eight were non-spinning waveforms with equal source frame component masses and a total mass of $76~M_{\odot}$; sixteen were aligned-spin with total masses from [70~M${}_{\odot}$, 80~M${}_{\odot}$] in the source frame and mass ratios from 1 to 5. Mass ratio is defined as $m_1 / m_2$ where $m_1 > m_2$. The waveforms were generated with the SEOBNRv2 approximant~\cite{seobnrv2}.

Since burst searches do not use gravitational-wave templates, they are less sensitive to compact merger signals than modeled searches~\cite{O1AllSky}. The burst searches did not detect GW151226, and recovered only a single hardware injection mentioned in Section~\ref{cbc} to validate that detection. The recovery of GW151226 hardware injections using the burst analyses are not discussed in this paper.

Coherent WaveBurst identifies coherent events in spectrographic data from the \aligo\ detectors constructed using a wavelet representation.
It then reconstructs the gravitational waveform signal using a contrained likelihood method~\cite{cwb}.
For signals consistent with compact binary coalescences, it also estimates the system's chirp mass based on the time-frequency evolution of the signal~\cite{Tiwari}.

\begin{figure*}       
\centering
\includegraphics[width=\columnwidth]{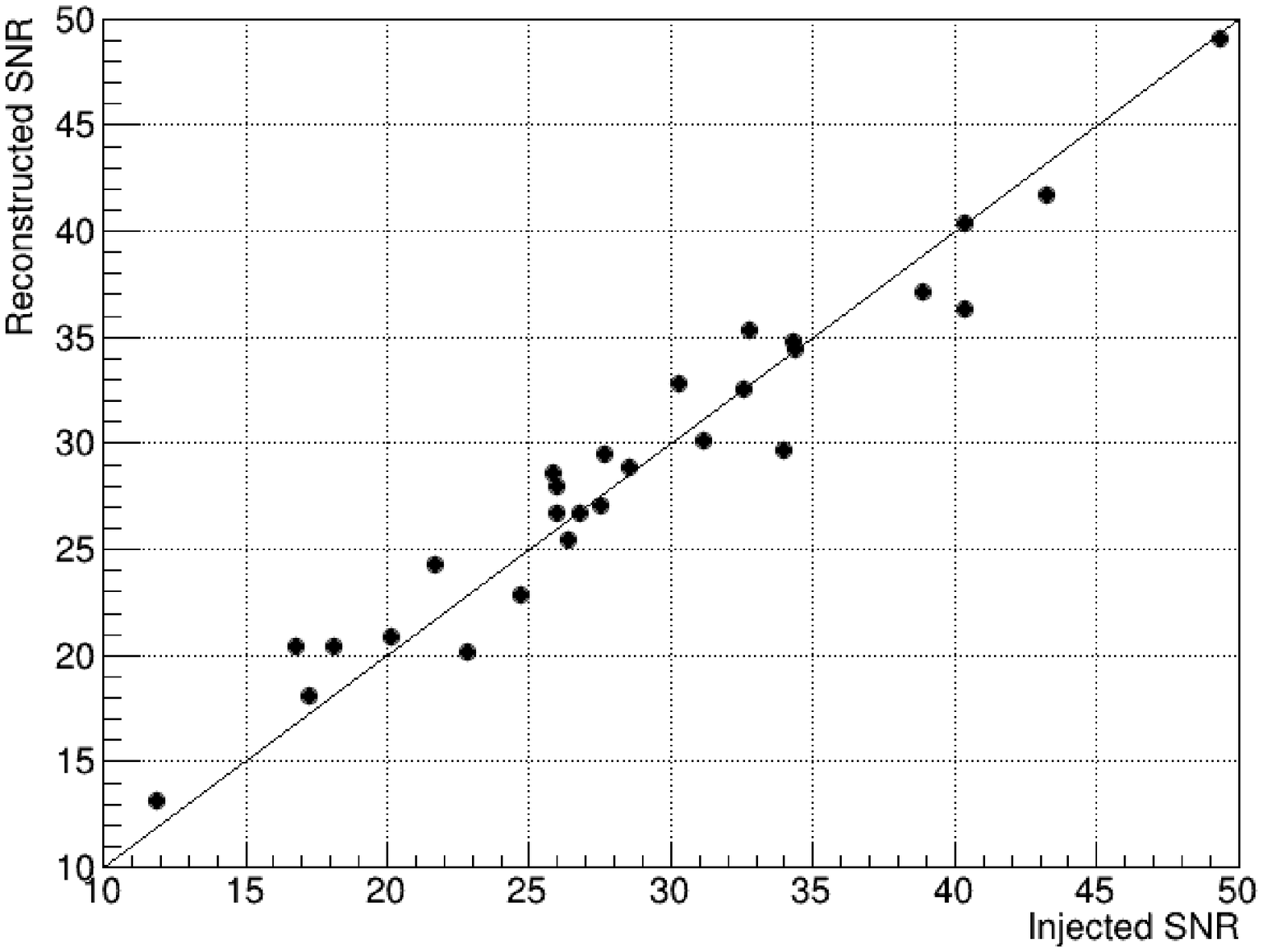}
\includegraphics[width=\columnwidth]{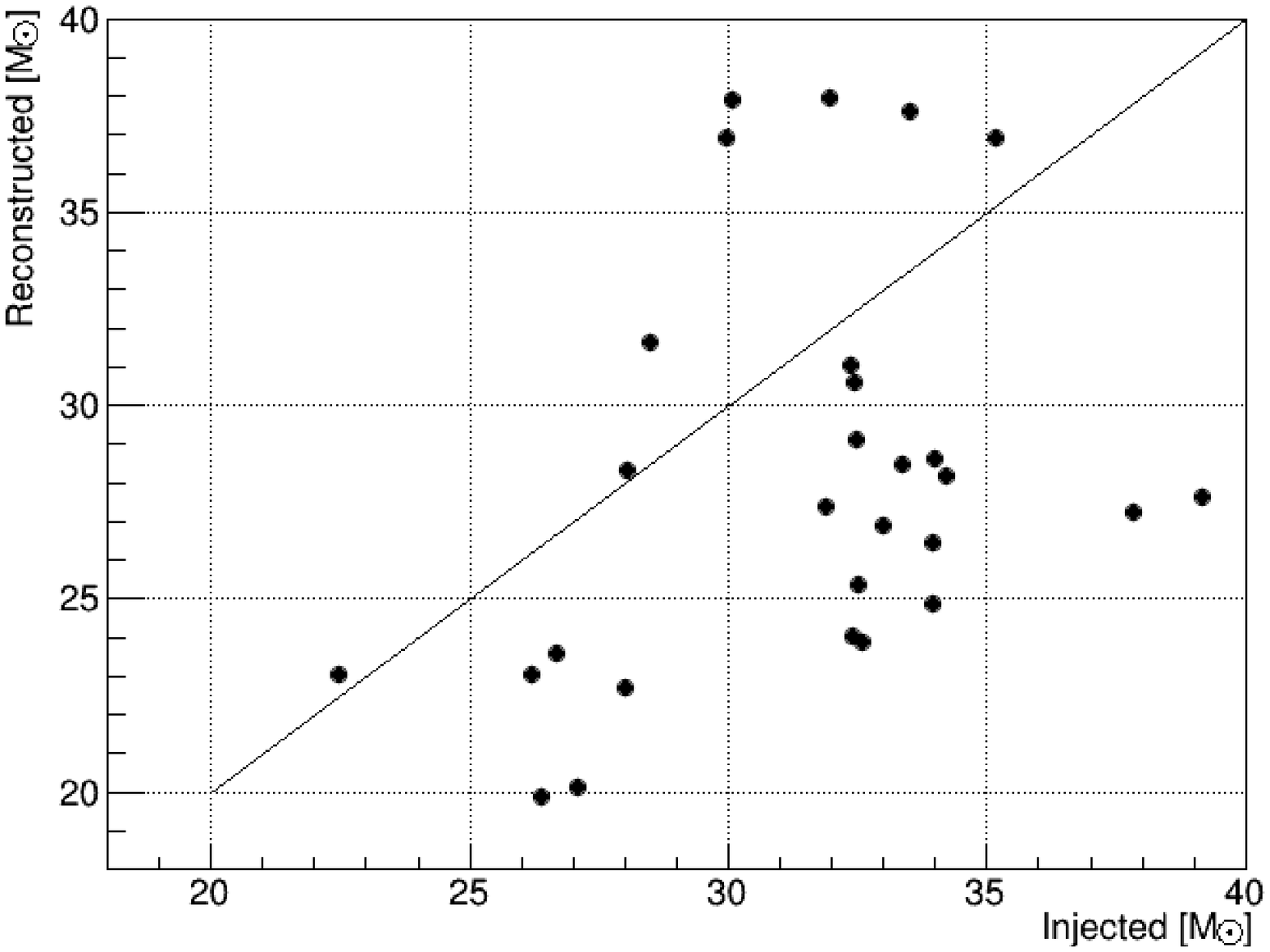}
\caption{
The 28 recovered hardware injections by the low-latency Coherent WaveBurst search. The recovered excess power signal-to-noise ratio and chirp mass are consistent with expectations~\cite{Tiwari}.
\emph{Left:} Recovered excess power signal-to-noise ratio (Reconstructed SNR) versus injected excess power signal-to-noise ratio (Injected SNR).
\emph{Right:} Recovered chirp mass (Reconstructed) versus injected chirp mass (Injected).
}
\label{inj_rec}
\end{figure*}

The low-latency Coherent WaveBurst search recovered 28 of the 34 total injections.
In Fig.~\ref{inj_rec} we show the recovered versus injected excess power signal-to-noise ratio and chirp mass.
Note that the excess power signal-to-noise ratio~\cite{cwb} is distinct from the matched-filter signal-to-noise ratio $\rho$.
Fig.~\ref{inj_rec} matches the expectations for the recovered chirp mass from fitting the time-frequency evolution of the data~\cite{cwb,Tiwari}.
These estimates help guide the initial response to detections. Subsequently, accurate estimates are obtained with template-based, fully coherent Bayesian parameter estimation~\cite{Veitch:2014wba}.

\begin{figure}
   \centering
   \includegraphics[width=\columnwidth]{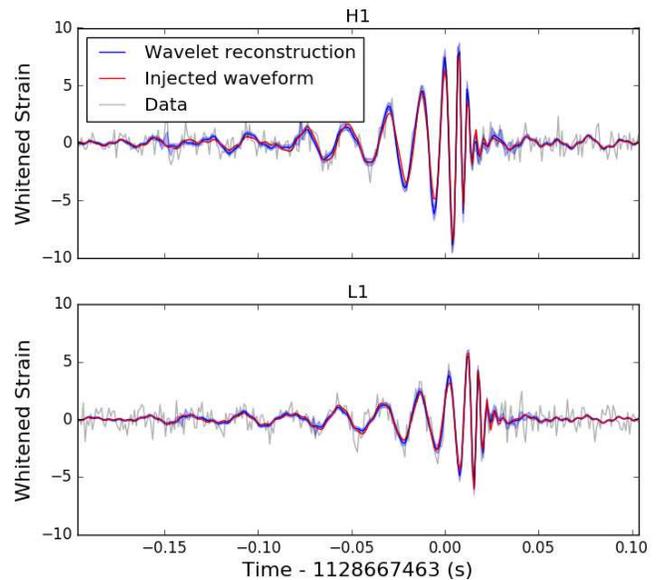} 
   \caption{
   BayesWave median reconstruction and 90\% credible interval (blue) and the injected waveform (red),
   time is shown on the $x$-axis and whitened strain on the $y$-axis.
   The data has been whitened using the estimated noise curve from the time of the injection.
   The network match for this waveform is 98\%.
   }
   \label{fig:BW}
\end{figure}

BayesWave uses a sum of sine-Gaussian wavelets to model the gravitational-wave signal~\cite{BayesWave}.
The reconstruction assumes an elliptically polarized gravitational wave, but no other constraints are imposed~\cite{BayesWave}.
BayesWave investigated the 28 GW150914 hardware injections recovered by Coherent WaveBurst.
Previous studies with software injections show that the recovered waveforms produced by BayesWave accurately match injected signals \cite{gw150914_burst}.
To measure the overlap between injected and recovered waveforms, we use the network match
\begin{equation}
\mathrm{Match}=\frac{(h_\mathrm{inj}|h_\mathrm{rec})}{\sqrt{(h_\mathrm{inj}|h_\mathrm{inj})~(h_\mathrm{rec}|h_\mathrm{rec})}}
\end{equation}
where $h_\mathrm{inj}$ is the injected waveform, $h_\mathrm{rec}$ is the recovered waveform, and $(a|b)$ is the noise-weighted inner product summed over all interferometers~\cite{BayesWave}.
The average network match between the injected and reconstructed waveforms is 94\%.
The 94\% match is consistent with the average match found using software injections \cite{gw150914_burst}.
An example of a reconstructed waveform is shown in Fig.~\ref{fig:BW}.

\begin{figure*}[ht]

\centering
\raisebox{-0.5\height}{\includegraphics[width=1.2\columnwidth]{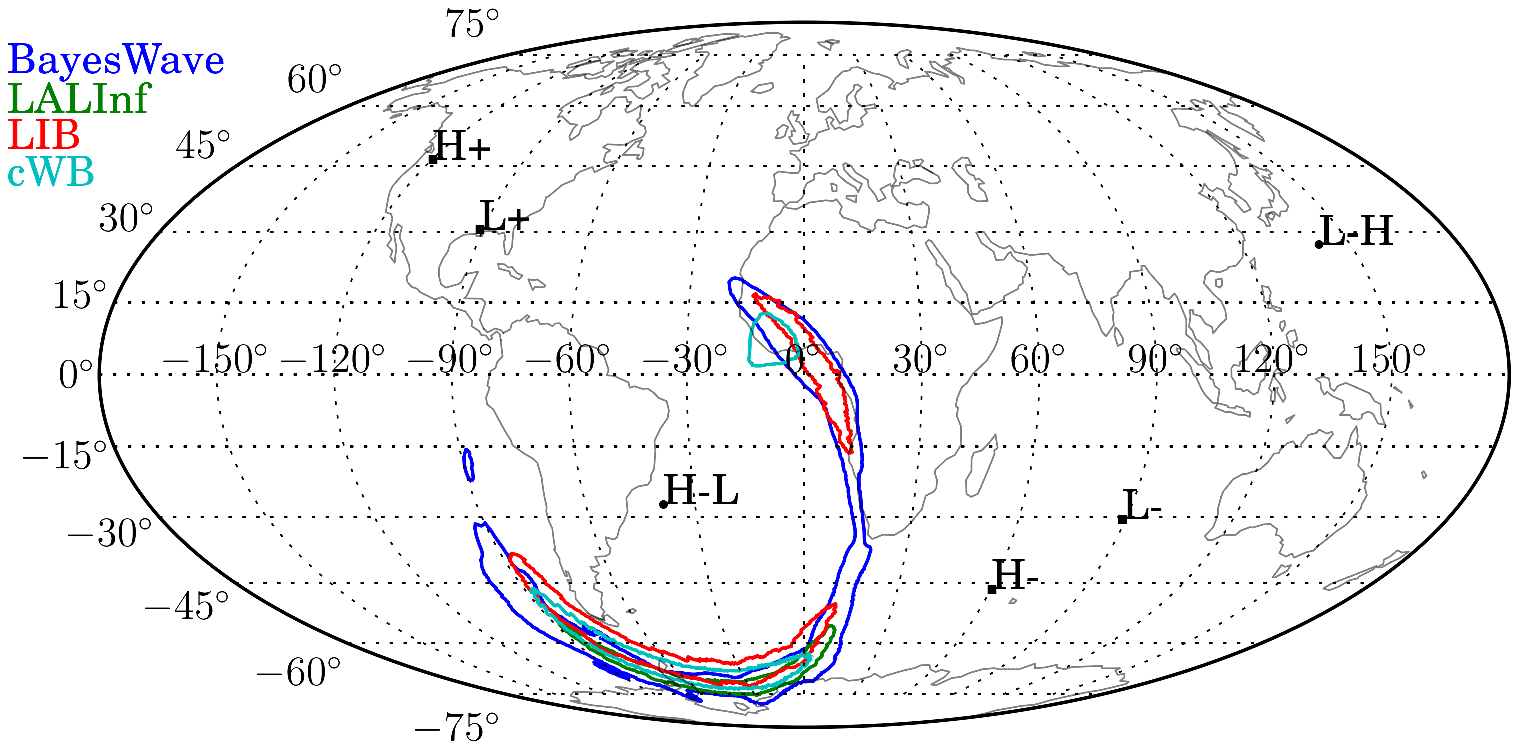}} \raisebox{-0.5\height}{\includegraphics[width=0.8\columnwidth]{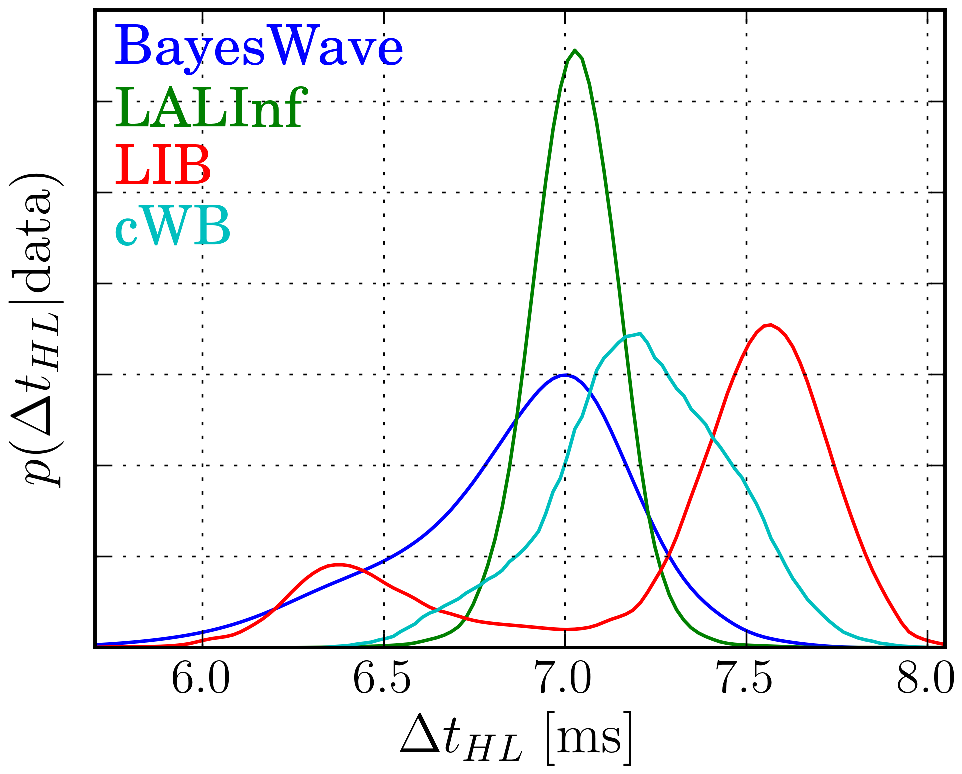}} \\

(a) Hardware injection at 1127783820 \\

\raisebox{-0.5\height}{\includegraphics[width=1.2\columnwidth]{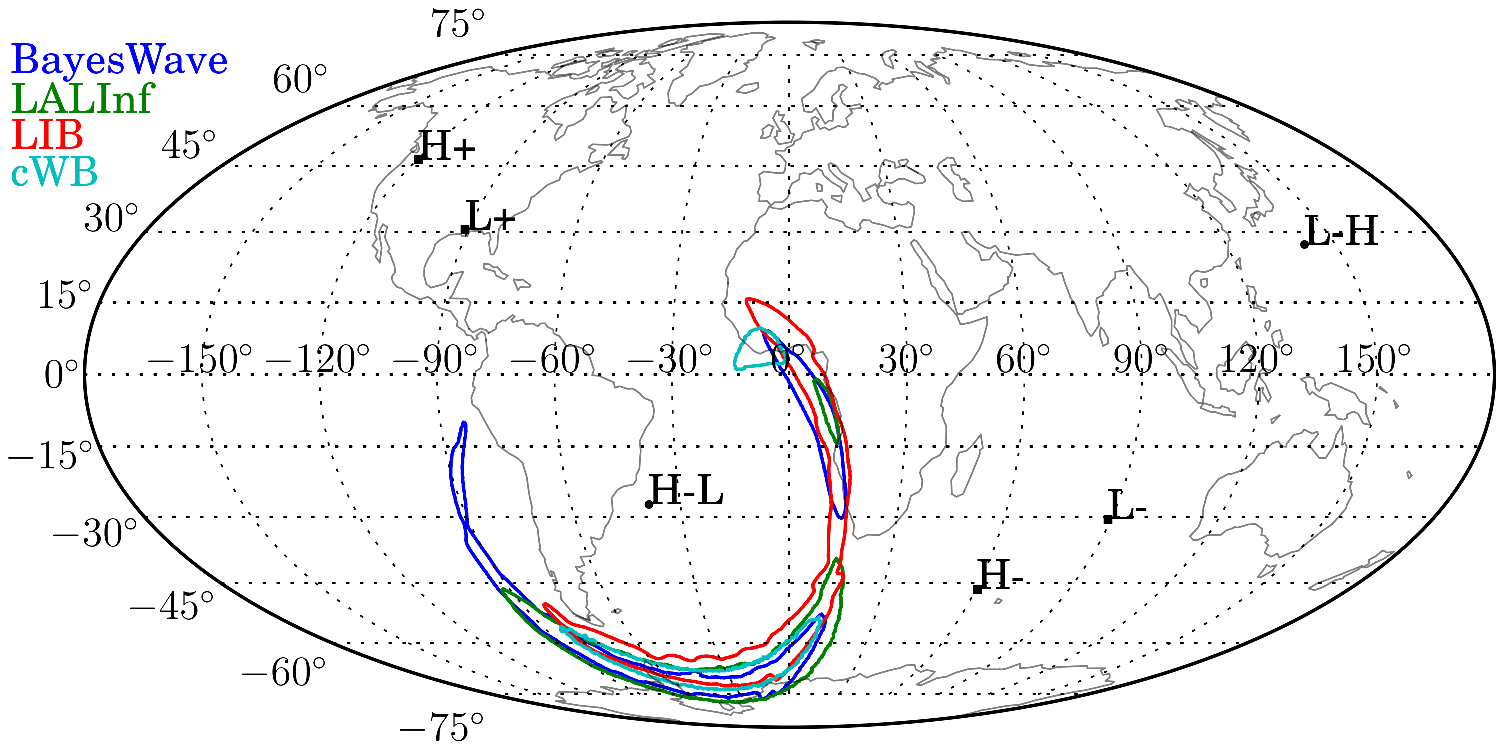}} \raisebox{-0.5\height}{\includegraphics[width=0.8\columnwidth]{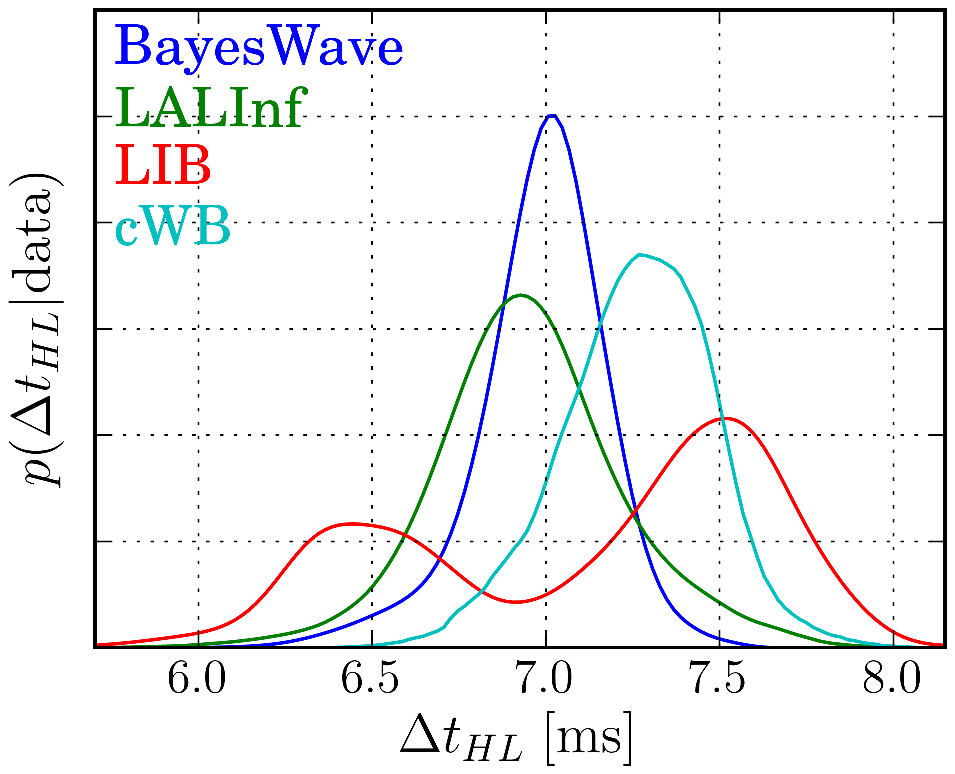}} \\
    (b) GW150914
  \caption{Sky localizations estimates for a GW150914 hardware injection and GW150914 itself from burst analyses: BayesWave, Coherent WaveBurst (cWB), and LALInferenceBurst (LIB). We include the parameter estimation analysis from Section~\ref{cbc} (LALInf) for comparison~\cite{gw150914_pe}. \emph{Left:} We show the localization maps in Earth-bound coordinates. \emph{Right:} To highlight the similar positions relative to the detectors the marginal distributions for the time delay between the two detectors is shown. Note that the horizontal scale is much smaller than the 10~ms light travel time between the two detectors.
  }
  \label{fig:burst_skymap}
\end{figure*}

Coherent WaveBurst, BayesWave, and LALInferenceBurst provide sky localization estimates. Fig.~\ref{fig:burst_skymap} demonstrates sky maps for one of the GW150914 hardware injections and GW150914 itself~\cite{gw150914_pe}. We see similar support in Earth-bound coordinates, and nearly identical structures around the triangulation rings. The right-hand panels of Fig.~\ref{fig:burst_skymap} highlight this with the posterior distributions for the time delay between the two detectors.

\subsection{Stochastic Gravitational-Wave \\ Hardware Injection}\label{sec:stochastic}
A stochastic gravitational-wave background is expected to arise from the superposition of many events, each of which are too weak to resolve, but which combine to create a low-level signal~\cite{2000PhR...331..283M}.
By cross-correlating data from two or more detectors, it is possible to detect low-level correlations hidden beneath the detectors' noise~\cite{1999PhRvD..59j2001A}.
The stochastic background from unresolved binary black holes is a particularly promising source, potentially within reach of advanced detectors~\cite{gw150914_stoch}.
For every binary black hole observed by \aligo, there are many more, which contribute to the stochastic background.
On October 23, 2015, a simulated stochastic gravitational-wave background signal was simultaneously injected into both detectors.
The $\unit[600]{s}$-long signal corresponded to one specific realization of an isotropic Gaussian background.
The background from binary black holes is actually non-Gaussian, but the standard stochastic search makes no distinction between Gaussian and non-Gaussian signals.

\begin{figure}[t]
\centering
\includegraphics[width=\columnwidth]{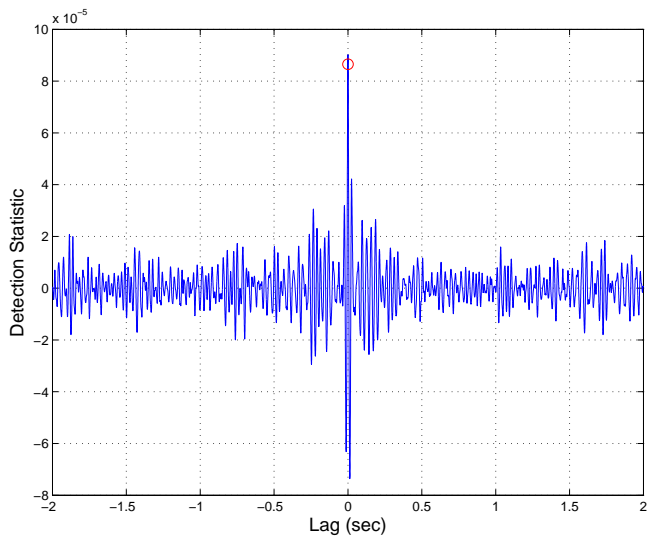}
\caption{
Recovered stochastic signal as a function of lag time between detectors (blue).
The detection statistic is an unbiased estimator for $\Omega(f)$ defined in Eq.~\ref{eqn:omega}.
The injected signal is represented with a red circle.
The peak of the recovered signal is at zero lag, indicating a successful recovery.
}
\label{fig:IFFT}
\end{figure}

\begin{figure}[t]
\centering
\includegraphics[width=\columnwidth]{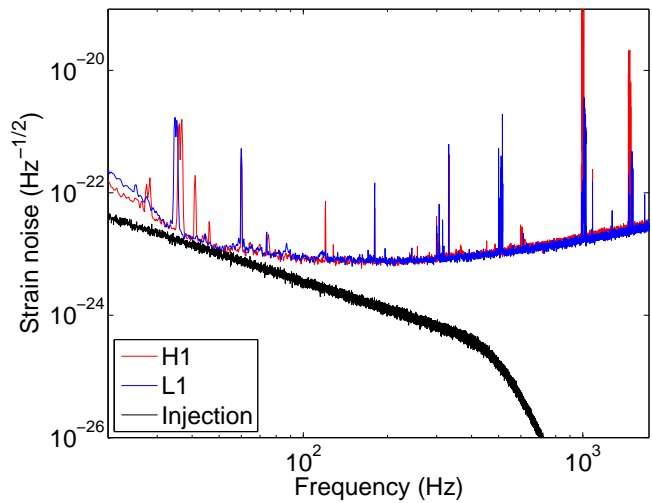}
\caption{
Amplitude spectral density of the injected stochastic signal compared to the instrumental noise at Hanford and Livingston.
}
\label{fig:inj_spectrum}
\end{figure}

The strength of a stochastic gravitational-wave signal is parameterized by the fractional contribution of the energy density in gravitational waves to close the Universe~\cite{1999PhRvD..59j2001A}:
\begin{equation}\label{eqn:omega}
  \Omega_{\rm GW}(f) = \frac{1}{\rho_c} \frac{\mathrm{d} \rho_\text{GW}}{\mathrm{d} \ln f} .
\end{equation}
Here, $\rho_c$ is the critical energy density of the Universe, $f$ is frequency, and $\mathrm{d} \rho_\text{GW}$ is the energy density between $f$ and $f + df$.
The injected signals were chosen such that $\Omega_{\rm GW}(f)=8.7\times10^{-5}$.
This corresponds to a strain power spectral density of 
\begin{equation}
  \begin{split}
    S_h(f)=& \frac{3H_0^2}{10	\pi^			2}\frac{\Omega_{\rm GW}(f)}{f^3}\\
    = &\left(\unit[2.9\times10^{-23}]{Hz^{-1/2}}\right)^2 \left(\frac{\unit[25]{Hz}}{f}\right)^3
  \end{split}
\end{equation}
where $H_0$ is the Hubble constant. Fig.~\ref{fig:inj_spectrum} shows the spectrum of the injected signal and the amplitude spectral density of the noise at Hanford and Livingston near the time of the injection. The signal was low-pass filtered below 500 Hz prior to injection.

We carried out a cross-correlation search following the standard procedure~\cite{s6vsr23_iso}.
The data was split into 50\%-overlapping, 60~s intervals, and utilizing coarse-grained 0.25~Hz-wide frequency bins, we recovered an $\Omega_\text{GW}(f)$ of $(8.8\pm0.6)\times10^{-5}$, consistent with the injected value.
Our injection recovery assumes that $\Omega(f)$ is constant.
The recovered signal (an unbiased estimator for $\Omega(f)$ in Eq.~\ref{eqn:omega}) is shown in Fig.~\ref{fig:IFFT}.
The $y$-axis shows the recovered signal as a function of the time lag between the detectors.
A peak at zero, and the absence of structure at other times, shows that the signal is recovered as expected.
\subsection{Continuous Gravitational-Wave \newline Hardware Injections}\label{sec:cw}

The recovery of hardware injections is used by continuous-wave searches as an end-to-end validation of the analyses in the presence of instrumental artifacts and imperfect instrument calibration.
Coherent searches for gravitational waves from known pulsars are sensitive to deviations from the injected signal since a small bandwidth around the gravitational-wave frequency is integrated for months or years~\cite{TargetedCWSearches}.
These searches have the capability of monitoring the self-consistency of the interferometer calibration and, in particular, the long-term stability of absolute phase recovery.
Continuous-wave searches can be implemented using a variety of methods, ranging from highly targeted searches based on ephemerides inferred from observed electromagnetic pulsations~\cite{TargetedCWSearches}, to systematic templated searches for sources with previously unknown frequencies and sky locations~\cite{S4PSH,S6PowerFlux,S6E@H}, to searches for excess radiation flux in narrow frequency bands~\cite{Ballmer:2005uw,s5searchdirectional}.
Here, we consider a coherent search based on Bayesian recovery of signal parameters~\cite{TargetedCWSearches} to validate the fidelity of hardware injections. This analysis can be used to cross-check elements of the instrument calibration, including proper coherence of data from interferometers separated by thousands of kilometers, which are sensitive to timing errors.

\begin{figure*}[t]
\centering
\subfigure{\includegraphics[width=\columnwidth]{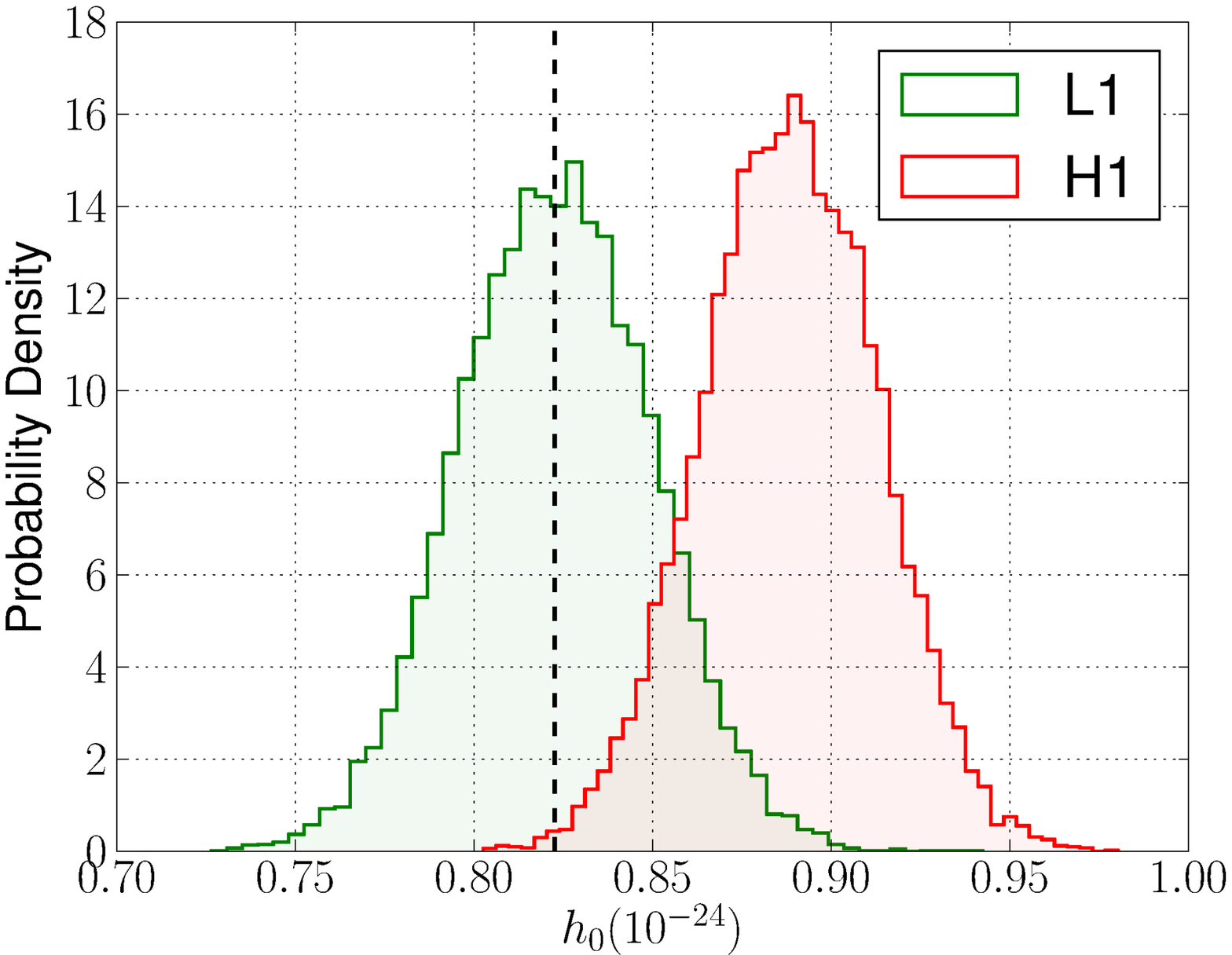}}
\subfigure{\label{fig:b}\includegraphics[width=\columnwidth]{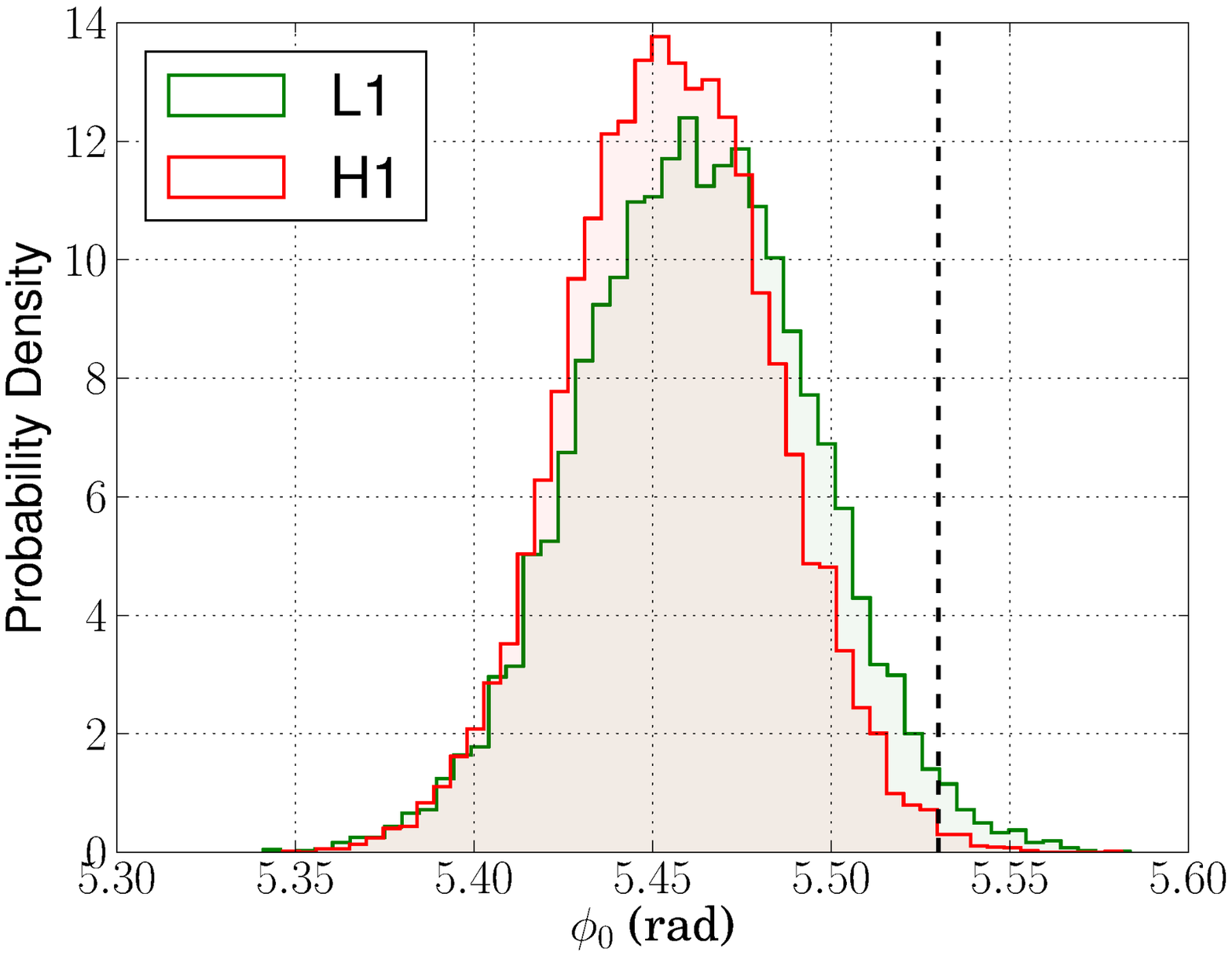}}
\caption{P03 posterior probability density functions for the recovered strain amplitude and phase constant for the injected spinning neutron star signal at $\unit[108.86]{Hz}$ (referred to as P03).
Note that the horizontal scales have suppressed zeroes.
The dashed vertical lines indicate the intended injection amplitude and phase in radians.
The red and green curves indicate the separately recovered amplitudes and phases for the Hanford and Livingston interferometers, respectively. The small discrepancies in
amplitude (10\%) and phase (0.1~radian = 5~degrees) fall within the uncertainties of the actuation system used for the injections.}
\label{fig:cw_pulsar3}
\end{figure*}

Fig.~\ref{fig:cw_pulsar3} shows the posterior probability density function (PDF) for strain amplitude $h_0$ for a simulated continuous sinusoidal gravitational waves emitted from a spinning neutron star, which has a signal frequency near 108.9 Hz and a nearly linear polarization.
The signal is recovered with an amplitude consistent with the intended strength, within calibration and actuation uncertainties.
Similarly, Fig.~\ref{fig:cw_pulsar3} also shows the recovered phase constant for this injection. In the analysis there is no compensation for the time delay between the hardware injection excitation channel and the detector output channel. The observed phase constant is consistent with the expected phase constant given the known (but uncompensated) time delay.

\begin{figure*}
\centering
\includegraphics[width=6in]{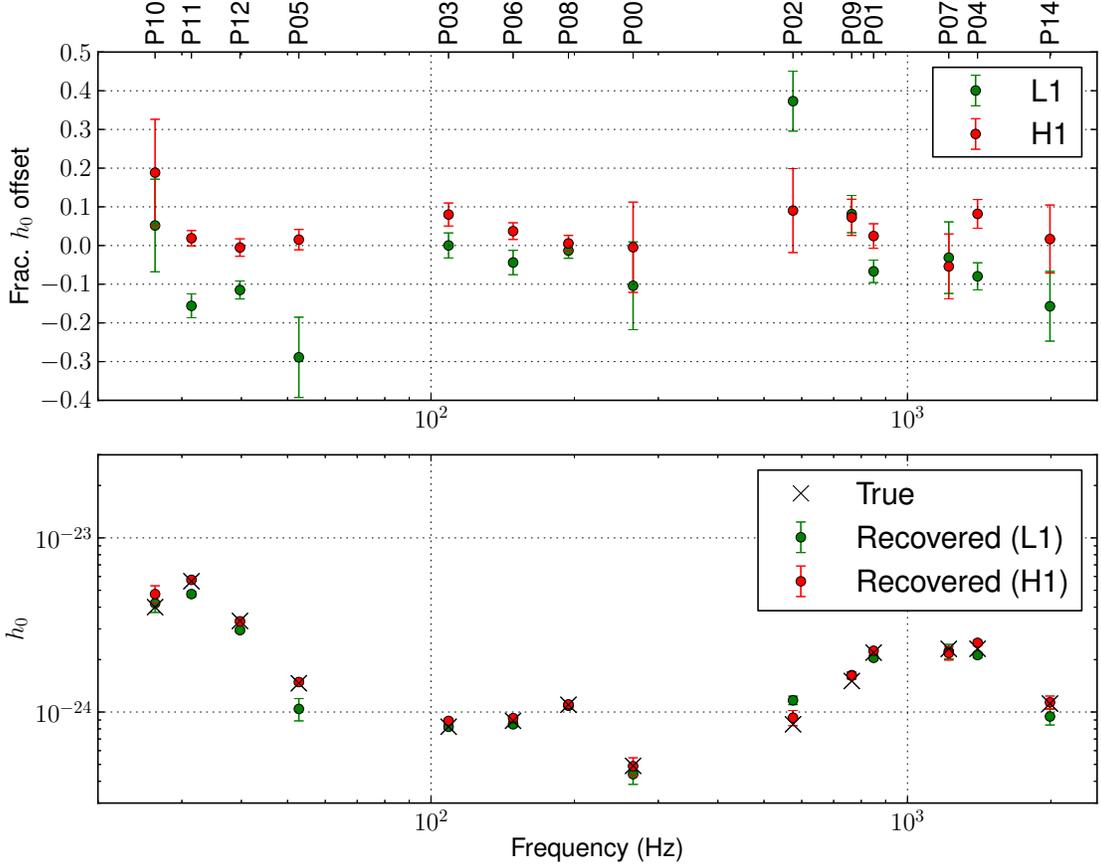}
\caption{
	Comparisons of recovered signal amplitudes for the Hanford and Livingston signals for the 14 recovered continuous-wave injections (P00-P14). 
    {\it 1st panel:} Fractional amplitude difference [(recovered minus injected)/injected]. {\it 2nd panel:} Amplitude values (recovered and injected).}
\label{fig:cwsummary_allpulsars_amps}
\end{figure*}

\begin{figure*}
\centering
\includegraphics[width=6in]{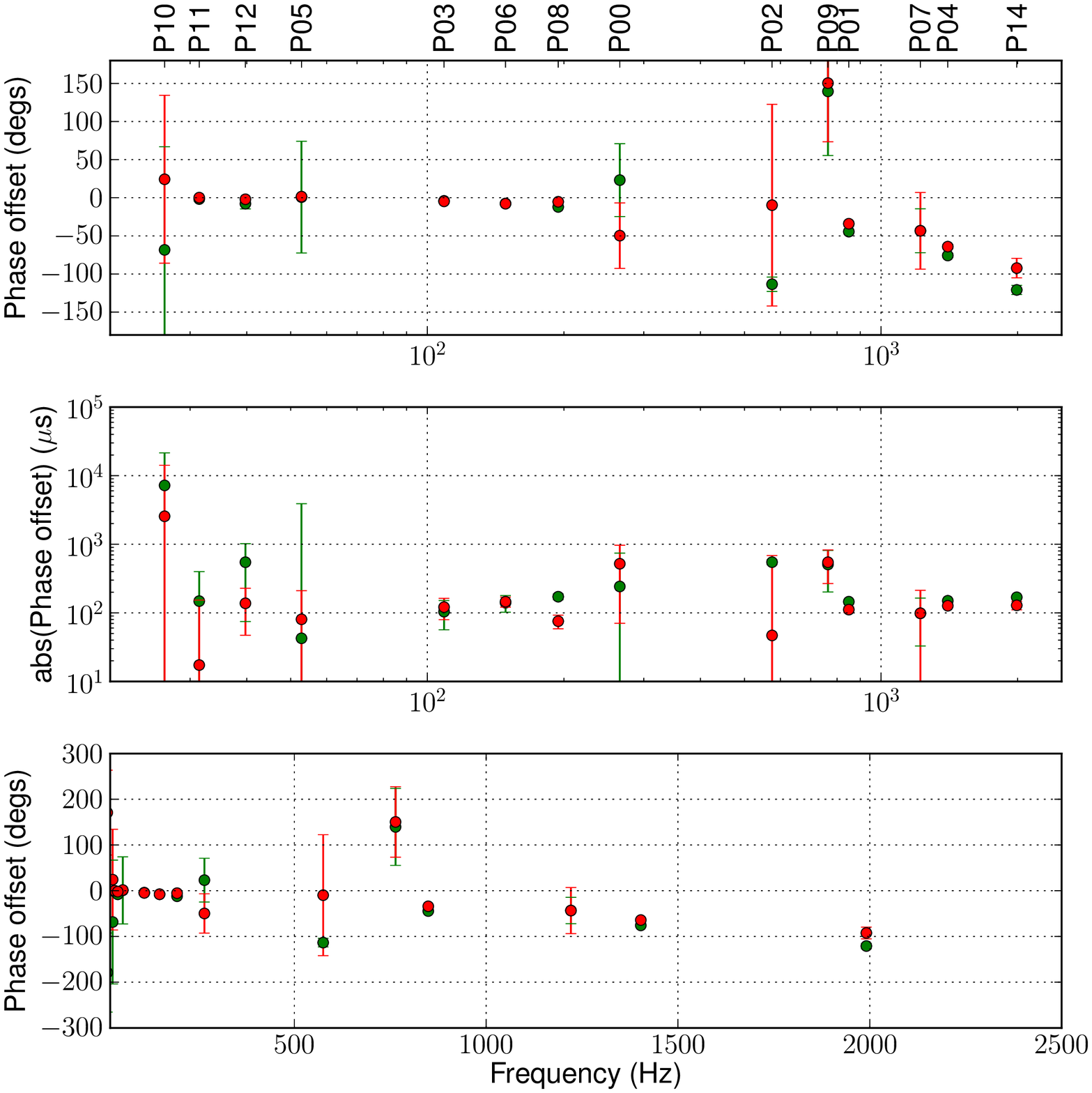}
\caption{
	Comparisons of recovered signal phases for the Hanford and Livingston signals for the 14 recovered continuous-wave injections (P00-P14). {\it 1st panel:} Phase difference (degrees) [recovered minus true]. {\it 2nd panel:} Conversion of phase difference magnitude to time difference. {\it 3rd panel:} Same as 3rd panel but with linear horizontal scale, to indicate the roughly linear dependence of residual phase difference, consistent with a constant known (but uncompenstated for) time delay.}
\label{fig:cwsummary_allpulsars_phases}
\end{figure*}

During \aligo's first observing run, the hardware injections included 15 simulated continuous gravitational wave as would be emitted by spinning neutron stars.
The signals were streamed in real-time with frequencies ranging from $\unit[12-1991]{Hz}$.
Figs.~\ref{fig:cwsummary_allpulsars_amps} and~\ref{fig:cwsummary_allpulsars_phases} show a summary of the agreement between recovered and intended amplitude and phase for the 14 injections with sufficient signal-to-noise ratio to permit recovery.
Instrumental noise at the lowest-frequency injection ($\unit[12]{Hz}$) proved too large were too large to permit recovery of simulated signal P13.
There is evidence of a constant uncompensated time delay of about $\unit[150]{\mu s}$ in the time-domain actuation, which manifests as a phase delay increasing linearly with injection frequency in Fig.~\ref{fig:cwsummary_allpulsars_phases}.
In the future, a compensating timing advance will be included in the inverse actuation filter, and the simulated continuous-wave sources' amplitudes will be increased for more precise and rapid validation of the hardware injection signal.

The overall sign of the calibration is important in order to detect and estimate the parameters of astrophysical signals correctly. An incorrect sign on the calibration would invert the signal in one detector and the parameter estimates would be incorrect.
The continuous-wave injections were used as an additional check on the sign of the calibration between the \aligo\ detectors, since an incorrect sign would lead to a relative phase offset between the two detectors in Fig.~\ref{fig:cwsummary_allpulsars_phases}. We found the sign of the calibration to be correct.

\subsection{Loud Hardware Injections \\ for Detector Characterization}\label{sec:detchar}

Noise artifacts in \aligo\ data adversely affect the output of gravitational-wave search analyses~\cite{DetcharGW150914,S6detchar}.
In searches for transient gravitational waves, some periods of time are excluded from the analysis to remove periods of poor data quality and known transient noise.
These are known as ``data quality vetoes''~\cite{DetcharGW150914,S6detchar}.
Removing periods of time with excess noise improves the performance of gravitational-wave searches~\cite{DetcharGW150914,S6detchar}.
Some of these data quality vetoes are derived from information recorded in auxiliary channels.
Auxiliary channels include instrumental channels that record degrees of freedom of the interferometer and its isolation systems as well as channels that monitor the environmental conditions around the instrument~\cite{effler2015environmental}. The environmental monitoring system includes seismic, acoustic, and electromagnetic data.

To avoid discarding true gravitational-wave signals, any auxiliary channels used for vetoes are first checked to ensure that they do not respond to gravitational-wave-like signals; i.e., changes in differential arm length. 
This process is referred to as a ``safety check,'' since a channel that has no sensitivity to gravitational waves is considered ``safe'' for use when constructing a veto.
To test whether auxiliary channels respond to differential arm length changes, three sets of 12 loud (matched-filter signal-to-noise ratios $>100$) transient hardware injections were performed at both detectors, and the auxiliary channel data were examined both qualitatively and quantitatively for signs of coupling. 

Spectrograms were manually inspected at the time of hardware injections.
These signals were very strong and clear, with high signal-to-noise ratio, in channels that were expected to record differential displacement, e.g.\ interferometer differential sensing and actuation, and closely related degrees of freedom.
No signs of coupling were found in thousands of other auxiliary channels, indicating that they may be used to construct vetoes.
Hundreds of time-frequency representations of auxiliary channels were also inspected at the times of GW150914 and GW151226 with the same outcome~\cite{DetcharGW150914}.  

Loud hardware injections were used to statistically assess the coupling.
An algorithm based on a transformation using sine-Gaussians~\cite{chatterji2004multiresolution} was used to identify and parameterize noise transients by their time, frequency, and signal-to-noise ratio.
The time of noise transient is compared with the times of the loud hardware injections.

For each channel, the number of noise transients that occurred within $\unit[100]{ms}$ of loud injections are counted and compared to the number that would be expected based on chance~\cite{hveto}.
For any channel exhibiting a higher number of overlaps than expected by chance, the time-frequency behavior of the raw data is further investigated to see if there is a plausible connection.
We find that only obviously related channels, such as those in the sensing and actuation chain for the differential length control loop, were sensitive to the loud hardware injections.

\section{Conclusions}\label{sec:conclusions}

This paper presents the \aligo\ hardware injection system infrastructure for injecting simulated gravitational-wave signals into the detectors by displacing the test masses, and results from \aligo's first observing run. Hardware injections were used for validating analyses after a gravitational-wave detection, as an additional check of the calibration, and characterizing the detectors' response to differential arm length variations.

After the detection of GW150914 and GW151226, sets of binary black hole merger waveforms with similar parameters were injected to validate the search and parameter estimation analyses.
The recovered signals were checked for consistency with the parameters of the injected waveforms, including signal-to-noise ratio, chirp mass, and sky position.
Similarly, the stochastic background and continuous-wave searches used simulated waveforms as an end-to-end test.

In order to detect and estimate the parameters of astrophysical signals the calibration must be correct, and the continuous-wave injections provided an additional check of the calibration sign.
They were also used to measure the time delay of the hardware injection pathway and checked that it was consistent with the predicted value from the calibration model.

Data quality vetoes are used to increase the performance of search analyses, and detector characterization hardware injections were used to identify output channels in the control system that can be used to construct data quality vetoes.
After each gravitational-wave detection, we carried out a study to check for cross-couplings with the detectors' output gravitational-wave strain channel. Channels that contained a trace of the injected signal were considered unsafe and excluded from data quality veto studies.

In the future, we plan to exclusively use the photon calibrators to inject simulated gravitational waves.
Future work on the hardware injection system includes using point-by-point, Fourier-domain inverse actuation functions for each of the injected spinning neutron stars to mitigate the effect of data dropouts.

\section*{ACKNOWLEDGEMENTS}

LIGO was constructed by the California Institute of Technology and Massachusetts Institute of Technology with funding from the National Science Foundation (NSF), and operates under cooperative agreement PHY-0757058. Advanced LIGO was built under award PHY-0823459.
Computations were carried out on the Syracuse University HTC Campus Grid which is supported by NSF award ACI-1341006.
Fellowship support from the LIGO Laboratory for S.~K. is gratefully acknowledged.
C.~B. and D.~A.~B. acknowledge support from NSF award PHY-1404395.
K.~R. acknowledges support from NSF award PHY-1505932.
E.~T. acknowledges support from the Australian Research Council award FT150100281.
P.~S. acknowledges support from NSF award PHY-1404121.
J.~R.~S. acknowledges support from NSF award PHY-1255650.
J.~V. acknowledges support from the Science and Technology Facilities Council award ST/K005014/1.
J.~L. and R. O. acknowledge support from NSF award PHY 1505629.
C.~B. would like to thank Laura Nuttall for providing useful suggestions and Collin Capano for the software injection data in Section~\ref{cbc}.
This paper carries the LIGO Document Number LIGO-P1600285.

\bibliography{bibliography}

\end{document}